\documentclass[12pt]{article}
\usepackage{epsfig}
\usepackage{hhline}

\begin{document}

\title{Multiscaling Anderson localization of cosmic 
electromagnetic fields}

\author{A. Bershadskii}

\maketitle
\begin{center}
{\it ICAR, P.O. Box 31155, Jerusalem 91000, Israel}
\end{center}

\begin{abstract}
Multiscaling properties of the Anderson localization of the cosmic 
electromagnetic fields before the recombination time are studied and 
results of a numerical simulation for a random banded matrix ensemble 
are found to be in good agreement with the large-scale cosmic data. 

\end{abstract}



\newpage 
\section{Introduction}

  Large-scale electromagnetic fields are expected to be generated 
by the cosmological phase transitions (electroweak and QCD) 
\cite{beo}-\cite{corn} before the recombination time. 
On the other hand, before the 
recombination time the cosmic plasma is strongly fluctuating at 
the large scales due to the gravitational density perturbation. 
These plasma density fluctuations can play a role 
of randomly distributed scatterers for the electromagnetic fields 
and make it impossible for the electromagnetic fields to propagate 
if the wave length is longer than the elastic mean free path of electromagnetic 
wave by the scatterers. The localized modes are thermally excited and 
the electric field is more or less aligned in the localization region. 
The localization phenomenon by random fields was 
first proposed by Anderson \cite{and} to explain the behavior of the 
electric resistivity of metals in the presence of impurities and the 
theories were developed by himself and his collaborators 
\cite{aalr},\cite{thou}. The localization phenomena were then observed in 
various physical systems. Probably the first paper which studied the possible 
localization of electromagnetic field in plasma is the Ref. \cite{es} 
(see also \cite{dvt}). 
Roughly speaking, the Anderson localization phenomenon occurs because 
of the destructive interference of the 
incident wave and the scattered waves from the randomly distributed 
sources. It is known that the localization occurs if the strength of 
the randomness is larger than a critical value. For a sufficiently 
large random potential, only the modes of the wave length shorter than 
the mean free path can propagate, while the other modes are localized. 
The authors of Ref. \cite{hk} argue that this localization phenomenon 
should occur in the matter dominated era (before the recombination time) 
in the standard cosmology. 
The authors of \cite{hk} took into account a single (Jeans) scale $l_c$ 
for the primordial perturbation for simplicity. Inclusion of a scale invariant 
spectrum for scales larger than $l_c$ produces a hierarchy of structures, which 
needs in a multifractal description. During the past decade, multifractality 
of critical eigenfunctions at the Anderson transition has been a subject of 
intensive analytical and numerical studies (see Refs. \cite{cp}-\cite{me} 
and references therein). A remarkable virtue of the Anderson localization 
mechanism is that the photon diffusion does not occur to the localized mode. 
This makes the localized structures to be imprinted in the photon gas after 
the decoupling of the photon gas (so-called cosmic microwave background, or 
CMB) from the baryonic matter. In the present paper we will show that results 
of the numerical simulations of multifractal Anderson localization are 
consistent with the CMB data. \\

\section{Anderson localization of cosmic electromagnetic fields}

Let us, following to the paper \cite{hk}, consider Anderson 
localization of cosmic electromagnetic fields for the spatially flat 
Friedmann-Robertson-Walker Universe. We assume that the charged 
particles are non-relativistic. So the four velocity of the charged 
particles: $u^{\alpha}=(a^{-1},v^i)$, with $v^i$ being small compared 
with unity. Equation for the vector potential ${\bf A}$  
in terms of the conformal comoving coordinates $(\lambda,{\bf x})$ in 
this case is 
$$
{\partial^2 {\bf A} \over \partial \lambda^2}+{\partial^2 {\bf A} 
\over \partial{\bf x}^2}
  ={ e^2na^2 \over m}{\bf A} \eqno{(1)}
$$
where $\lambda$ is the conformal time defined by 
$\lambda= \int^t{dt \over a}$, $m$ is the electron mass and $e$ 
is the charge, $n$ is the density.

Split the plasma density $n$ into the homogeneous part $\overline{n}$ 
and the space dependent random part $\delta n$; 
$$
n = \overline{n}+\delta n  \eqno{(2)}
$$
Then the right hand side of the plasma equation becomes 
$$
[{\overline{\omega_p^{rec}}^2 \over a} +{\overline{\omega_p^{rec}}^2 
\over a} {\delta n \over \overline{n}}]{\bf A}  \eqno{(3)}
$$
where $\overline{\omega_p^{rec}}=\sqrt{{e^2 n_{rec} \over m_e}}$. 
Here we have used the conservation law: $\overline{n}=n_{rec}/a^3$ 
before the recombination time, with the scale factor $a$ normalized 
at the recombination time. $n_{rec}$ is the electron density 
(therefore the baryon density) at the recombination time. 
The first term in the bracket is spatially constant but a 
decreasing function of the conformal time $\lambda$. 
The second term represents the fluctuating random 
environment induced by the baryonic density perturbation 
${\delta\rho_b \over \rho_b}$ which is equal to 
${\delta n \over \overline{n}}$ in our non-relativistic case. 
Knowing the standard theory of linear perturbation in cosmology 
we see that ${\delta n \over \overline{n}}={\delta\rho_b \over 
\rho_b} \propto a$ in the matter dominated era, so that the 
second term is independent of the conformal time in that era. 
We write the random potential as $V({\bf x})\equiv 
{\overline{\omega_p^{rec}}^2 \over a}{\delta n \over \overline{n}}$, 
which is a function of comoving coordinates ${\bf x}$ only. 
So far we have discussed the period after the equal 
time\footnote{The equal time 
$t_{eq}$ is the time when the matter and radiation energy 
density are equal. This value depends on the value of the 
present baryon density. We take ${a(t_{rec})
\over  a(t_{eq})}= 
({t_{rec} \over t_{eq}})^{2/3}=6$ as a typical value for 
$\Omega_{0}=0.2$.} 
and before the recombination time and found that the random 
potential induced by the density perturbation is constant in time. 
As we shall see this is important for the stability of the 
Anderson localization. 
For the matter dominated universe, the scale factor is given by 
$a(t) =({t \over t_{rec}})^{2/3}=({\lambda \over 3t_{rec}})^2 $ 
with $\lambda =\int{dt \over a(t)}=t_{rec}^{2/3}3t^{1/3}$. 
One can reduce the Maxwell equations 
to the following wave equation as 
$$
{\partial^2 {\bf A} \over \partial \lambda^2}+
{\partial^2 {\bf A} \over \partial{\bf x}^2}= 
[Q^2/\lambda^2+V({\bf x})]{\bf A}  \eqno{(4)}
$$
Here $Q^2=9t_{rec}^2\overline{\omega_p^{rec}}^2$. 
We expand the vector potential ${\bf A}$ in terms 
of the complete orthonormal set $\{\psi_n\}$ as 
$$
{\bf A}=\sum {\bf a}_nf_n(\lambda)\psi_n({\bf x})+c.c.  
\eqno{(5)}
$$
with ${\bf a}_n$ being arbitrary coefficients. 
Here $\psi_{n}({\bf x})$ is a normalized solution of the eigenvalue 
equation:
$$
[-\Delta+V({\bf x})]\psi_n({\bf x})=E_n^2\psi_n({\bf x}) \eqno{(6)}
$$

Localization means that a number of modes $\{\psi_n, n\leq n_0\}$ 
are bound states exponentially falling off at infinity. 
The marginal state $ n_0$ corresponds to
the so-called mobility edge as we shall explain further. \\

Here we only consider the adiabatic baryonic 
density perturbation for simplicity, the spectrum of which is 
roughly like the picture below. To simplify our discussion we 
assume in this section that the spectrum is 
sharply peaked around the Jeans 
length $l_c$ at the equal time and the distribution of 
perturbation is Gaussian. Inclusion of the scale invariant 
spectrum for scales larger than $l_c$ produces a hierarchy of 
structures (see next section).

To understand the Anderson localization in the Universe 
before the recombination time, we have to evaluate the mobility 
edge of the wave number, which is roughly the inverse of 
the mean free path $l_{mf}$ of the electromagnetic wave by the 
random potential. Consider a Schr\"{o}dinger equation for the 
spatial part of the electromagnetic field: 
$$
[-\Delta+V({\bf x})]\psi=E^2\psi  \eqno{(7)}
$$
with ${\bf x}$ being the comoving coordinates. 
Recall that the random potential $V({\bf x})=
\overline{\omega_p^{rec}}^2{\delta n \over \overline{n}}$ 
does not depend on the conformal time in the matter dominated era. 
The characteristic length scale $l_c$ of the density perturbation 
is given by the Jeans length at the equal time, which is roughly 
the same as the horizon 
scale at that time. After that time the
length scale $l_c$ 
develops as $\propto a$. This gives $l_c\approx10^{23}cm$ at the 
recombination time, while the horizon is $\approx 2\times10^{23}cm$. 
Then the mean free path is given by 
$l_{mf}={1 \over n_c\sigma} $
where $\sigma$ is the elastic scattering cross section of 
the electromagnetic wave by a single potential and 
$n_c\approx{3 \over 4\pi} l_c^{-3}$ is the density of 
"impurities". The elastic cross section is estimated roughly 
as $\approx 4\pi l_c^2$ by the standard wave mechanics, 
because the "energy" is low and is roughly given by 
$E^2 l_c^2\approx 1$, while the potential is 
effectively gigantic $V\times l_c^2\approx 
\overline{\omega_p}^2{\delta n \over n}l_c^2\approx 5\times10^{32}$. 
Here we have taken ${\delta n \over n}\approx10^{-5}$ 
because the assumed baryonic perturbation can be comparable 
to the temperature fluctuation of CMB observed by COBE (see 
\cite{COBE} and next section). 
However, the localization length and other results in the present 
consideration are insensitive to the magnitude of the density perturbation. 
Therefore the localization should occur, since the critical value 
of $Vl_c^2$ will be of order one. Then the mean free path is 
given by $l_{mf}\approx l_c/3$ 
and the mobility edge is $E^{*}\approx {12\pi \over l_c}$, 
which is consistent with our low energy picture. \\

Fortunately in the present case, the randomness is huge so 
that an estimate of the localization length is available 
\cite{am}. We have $\xi=l_{c}/\log(|V|/E^2)$, 
where $l_{c}$ is the minimum length scale 
in which range the potential can be considered as a constant. 
For the adiabatic baryonic perturbation, it will be reasonable 
to take the Jeans length at the equal time $t_{eq}$ 
as the small scale cut-off $l_{c}$, which is $\approx 10^{23} cm$ 
at the recombination time. In the present case, 
$|V|/E^2\approx5\times10^{32}$ so that $\xi\approx 10^{21} cm$ 
at the recombination time. \\

   One can estimate the magnitude of the 
thermally excited localized electric fields as 
$\overline{{\bf E}_{loc}^2}\approx T/\xi^3$
in a localization region of the size $\xi$. 
Here the contribution from a single 
localized mode only has been taken into account, 
because the peaks of the other localized modes are 
randomly distributed and will be 
located outside of the region under consideration. 
The estimate of $\overline{{\bf E}_{loc}^2}$ can be 
derived, for instance, by the equal partition law. 
Note that the propagating modes($n>n_0$) contribute 
to the Stefan-Boltzmann law. At this stage we realize 
that approximately $T/\omega_p$ photons occupies
a localized state. As far as $T/\omega_p>>1$, a coherent 
localized state of the size $\xi$ is a good picture so that we can 
approximately describe it by a classical
field: 
$$
{\bf E}_{loc}= -\sum ^{n_0}\sqrt{T/\omega_p}\dot{f_n}(\lambda)
{\bf e}_n({\bf x})\psi_n({\bf x})+c.c.  \eqno{(8)}
$$
where ${\bf e}_n({\bf x})$ is the polarization vector, 
which is perpendicular to the gradient of the wave function 
$\psi_n({\bf x})$. Therefore, in a localization domain of the size 
$\xi$ in the cosmic plasma, we have aligned electric
and magnetic fields, which are perpendicular to each other and 
oscillate almost at the plasma frequency. 
One might worry about the damping of the electromagnetic structure of the 
size $\xi$ by photon diffusion. However, the diffusion does not occur 
to the localized modes. This is one of the virtues of the 
Anderson localization mechanism. 
Since the localization is a delicate interference phenomenon, 
the random potential $V({\bf x})$ should vary slowly so that 
the motion of wave
packet is not disturbed \cite{es}. This 
condition is met before the recombination time. 
In the previous consideration we, following to \cite{hk}, only took 
into account a single scale $l_c$ for the primordial perturbation 
for simplicity. The contribution from the larger scales to 
the localization phenomenon is interesting in the sense that 
it may account for the hierarchy of the structures. 

\section{Multifractal Anderson localization}

  Even arbitrary close to the mobility edge a state should 
occupy only an infinitesimal fraction of space if it is to 
be labeled a localized state. On the other side of the 
mobility edge the states should extend through the whole 
volume. Both characteristics can be accommodated at the 
mobility edge if one assumes a fractal wave function with 
filamentary structure likes a net over the whole volume, as 
suggested originally by Aoki \cite{aoki}. Observations 
of the scaling properties of different characteristics of the 
eigenstates show that the wave functions cannot be adequately 
treated as a simple fractal. Rather, the more general concept 
of multifractality has to be employed, yielding a set of 
generalized dimensions. 

  The multifractal fluctuations of eigenfunctions can be 
characterized by a set of inverse participation rations
$$
P_p = \int d^d r |\psi ({\bf r})|^{2p}  \eqno{(9)}
$$
with an anomalous scaling with respect to system size $L$
$$
P_p \sim L^{-\tau_p}, ~~~~~~~~~\tau_p=D_p(p-1)   \eqno{(10)}
$$
 The scaling (10) characterized by an infinite set  of 
generalized dimensions $D_p$ implies that the critical eigenfunction 
represents a multifractal distribution. During the past decade, 
multifractality of critical eigenfunctions at the Anderson transition 
has been a subject of intensive numerical studies see Refs. 
\cite{cp}-\cite{me} and references therein. In recent paper 
\cite{me} the authors explore the fluctuations at criticality 
for the power-law random banded matrix (PRBM) ensemble. The 
model is defined as the ensemble of random Hermitian $N$x$N$ matrices. 
The matrix elements $H_{ij}$ are independently distributed Gaussian 
variables with zero mean $\langle H_{ij} \rangle =0$ and the variance 
$$
\langle |H_{ij}|^2 \rangle = a^2 (|i-j)|),  \eqno{(11)}
$$
where $a (r)$ is given by
$$
a^2(r) =\frac{1}{1+(r/b)^{2\alpha}}   \eqno{(12)}
$$
At $\alpha=1$ the model undergoes an Anderson transition from the 
localized ($\alpha >1$) to the delocalized ($\alpha <1$) phase. 
At $\alpha =1$ the PRBM model was found to be critical for 
arbitrary value of $b$; it shows all the key features of the Anderson 
critical point, including multifractality of eigenfunctions.  
In a straightforward interpretation, the PRBM model describes an 
one-dimensional sample with random long-range hopping, 
the hopping amplitude decaying as $1/r^{\alpha}$ with the 
length of the hop. 
Such a random matrix ensemble arises in various contexts
in the theory of quantum chaos \cite{js},\cite{al} and disordered systems \cite{bs},\cite{ps},\cite{skf}.
 In the paper \cite{me} the authors studied 
the periodic (critical) generalization of (12)
$$
a^{-2}(r)=1+\frac{1}{b^2} \frac{sin^2 (\pi r/N)}{(\pi /N)^2}  
\eqno{(13)}
$$
Indeed, for $N \gg r$ Eq. (13) reduces to Eq. (12) at the critical 
case. The generalization 
(13) allows diminish finite-size effects (an analog of periodic boundary 
conditions). At $b=1$ a limiting distribution can be reached most easily 
and, therefore, the finite-size effects are minimal \cite{me}. This value 
of parameter $b$ corresponds to crossover from weak ($b \gg 1$) to strong 
($1 \gg b$) multifractality. Below we will compare results of the numerical 
simulations performed for $b=1$ with the large-scale CMB data.\\ 

Temperature ($T$) dissipation rate of the photon (CMB) gas can be 
characterized by a "gradient" space measure (\cite{my}, p. 381):
$$
\chi_r =\frac{\int_{v_r} (\bigtriangledown{T})^2 dv}{v_r}
    \eqno{(14)}
$$
where $v_r$ is a subvolume with space-scale $r$. 
Scaling law of this measure moments,
$$
\frac{\langle \chi_{r}^p \rangle}{\langle \chi_{r} \rangle^p} 
\sim r^{-\mu_p}      \eqno{(15)}
$$
(where $\langle ... \rangle$ means an average) is an 
important characteristic of the dissipation rate \cite{my},\cite{s}. 
The exponents $\mu_p$ can be related to correspondent generalized 
dimensions by equation \cite{s}:
$$
\mu_p =(d-D_p)(p-1)   \eqno{(16})
$$ 

The COBE satellite was launched on 1989 into 900 km 
altitude sun-synchronous orbit. The Differential Microwave 
Radiometers (DMR) operated for four years of 
the COBE mission and mapped the full sky. 
The instrument consists of six differential microwave 
radiometers, two nearly independent channels that operate at each of 
three frequencies: 31.5, 53 and 90 GHz. Each differential radiometer 
measures the difference in power received from two directions in the 
sky separated by 60$^0$, using a pair of horn antennas. 
Each antenna has a 7$^0$ beam. In this investigation we use 
a DMR data map of the cosmic microwave radiation temperature. 
The COBE-DMR team made an effort to remove the most of the Galactic 
microwave emission (i.e. thermal emission of cosmic dust, 
free-free emission and synchrotron emission) 
from the map using so-called "combination technique" in which a 
linear combination of the DMR maps is used to cancel the Galactic 
emission. This was made in order to extract the cosmic microwave 
background radiation. 
 
To organize the DMR data in the temperature maps the sky was 
divided into 6144 equal area pixels. These 
are formed by constructing a cube with each face 
divided into 32 x 32 = 1024 squares, projected onto a celestial 
sphere in elliptic coordinates. The projection 
is adjusted to form equal area pixels having a solid angle of 
4$\pi$/6144 sr or 6.7 square degrees. Because the $7^0$ beamwidth of 
the sky horn is greater than separation between pixels ($2.6^0$ average), 
this binning oversamples the sky. 

Using the CMB pixel map, we will 
calculate the CMB temperature gradient measure using 
summation over pixel sets instead of integration over 
subvolumes $v_r$. So that multiscaling (15) (if exists) 
will be written as
$$
\frac{\langle \chi_s^p \rangle}{\langle \chi_s \rangle^p} 
\sim s^{-\mu_p}      \eqno{(17)}
$$
where metric scale $r$ (from (15)) is replaced by 
number of the pixels, $s$, characterizing size of the 
summation set. The $\chi_s$ is a one-dimensional 
surrogate of the real dissipation rate $\chi_r$. 
It is believed that the surrogates 
can reproduce quantitative multiscaling properties of 
the dissipation rate \cite{s}. Since in our case 
$\langle \chi_s \rangle$ is independent on $s$, we will 
calculate the exponents $\mu_p$ directly from scaling of 
$\langle \chi_s^p \rangle$.\\

Figure 1 shows scaling of the CMB temperature 
dissipation rate moments $\langle \chi_s^p \rangle$ 
calculated for the DMR map. The straight lines 
(the best fit) are drawn to indicate 
the scaling in the log-log scales. 

Figure 2 shows the generalized dimensions $D_p$ extracted 
from figure 1 (circles) using Eq. (16). Crosses in this figure 
correspond to the generalized dimensions calculated 
for the above described PRBM model \cite{me} of the Anderson 
localization. 

\section{Discussion}  
 
    The Anderson localization is an unique {\it linear} 
phenomenon which exhibits profound multifractality. This allows 
explain the observed multifractality of the CMB dissipation 
in the frames of standard (linear) approach to the CMB, which 
seems to be the most compatible with the small magnitude of the 
CMB fluctuations. Generation of the cosmic electromagnetic 
fields in the electroweak or QCD phase transitions occurring 
in the particle physics scales needs in a non-linear (turbulent) mechanism for reasonable space scale expansion 
of the fields \cite{beo}-\cite{corn}. Then the Anderson 
localization of the large-scale electromagnetic fields 
can be considered as an effective linear mechanism of the final 
CMB fractalization. This is because the photon 
diffusion, which can effectively suppress the non-linear 
fluctuations just before the recombination time, does not occur 
to the localized modes. \\

The author is grateful to C.H. Gibson and to K.R. Sreenivasan 
for discussions, and to A.D. Mirlin, to F. Evers and 
to the NASA Goddard Space Flight Center for providing the data.

\newpage

\centerline{\bf Figure Captions}

Figure 1.  Scaling of the CMB temperature 
dissipation rate moments $\langle \chi_s^p \rangle$ 
calculated for the four-year DMR map. The straight lines 
(the best fit) are drawn to indicate 
the scaling in the log-log scales.\\

Figure 2. Generalized dimensions $D_p$ extracted 
from figure 1 (circles) using Eq. (16). The crosses 
correspond to the generalized dimensions calculated 
for the PRBM model \cite{me} of the Anderson 
localization.

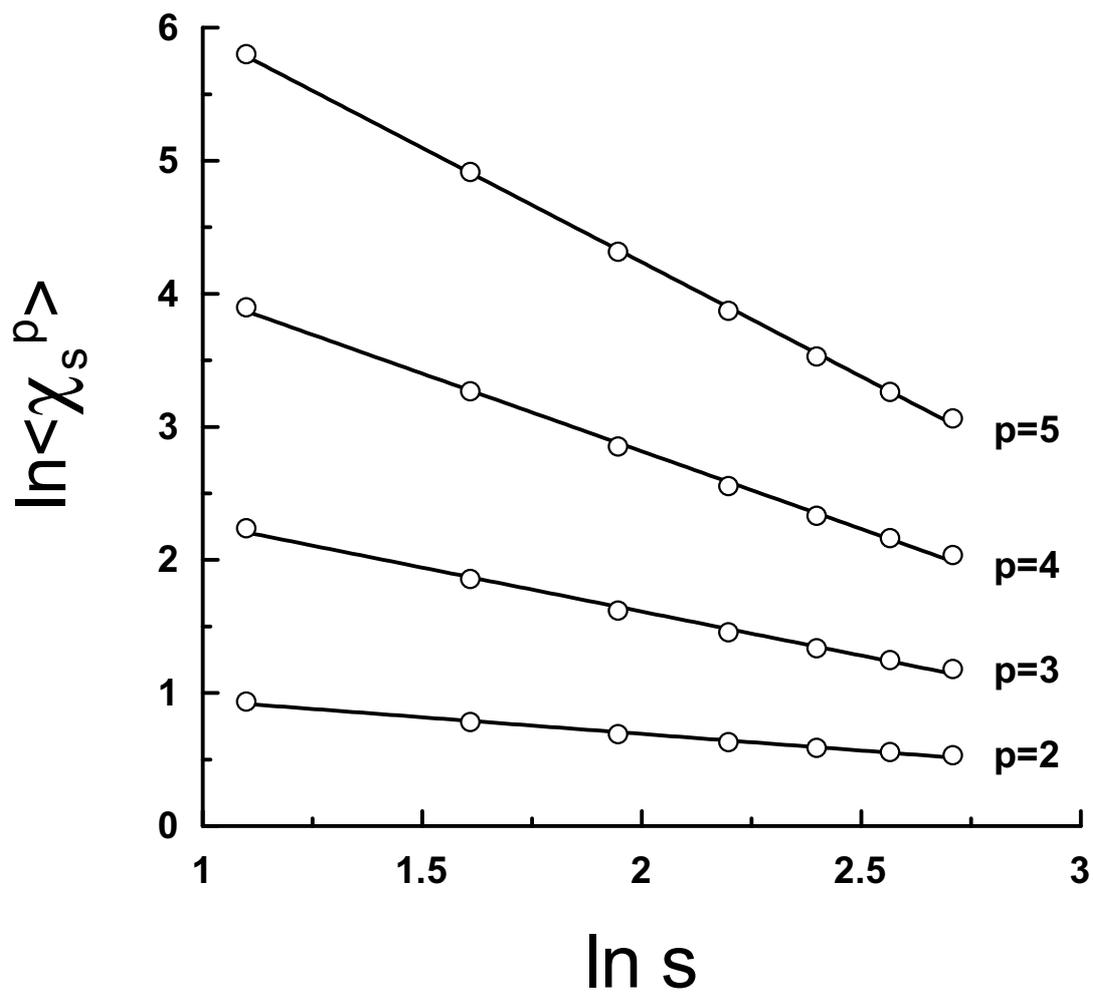
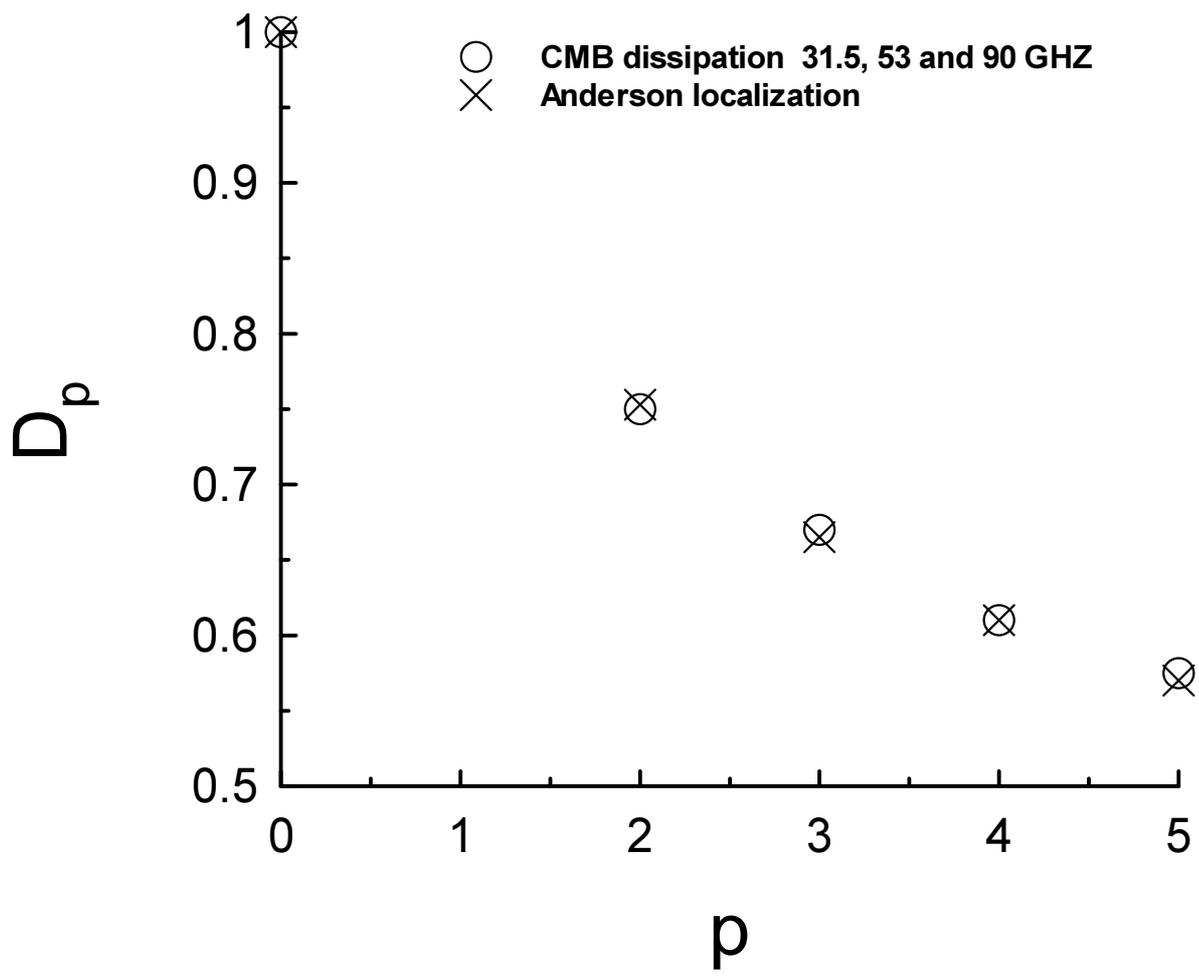
\end{document}